\documentclass[journal,10]{IEEEtran}
\usepackage[latin1]{inputenc}
\usepackage{amssymb}
\usepackage[dvips,pdftex]{graphicx}
\usepackage{amssymb}
\usepackage{caption}
\usepackage{subcaption}

\usepackage{verbatim}
\usepackage[normalem]{ulem}
\usepackage{lineno}
\usepackage{setspace}
\usepackage{color}
\usepackage{url}
\usepackage{cite}
\usepackage{footnote}

\usepackage{float}
\usepackage{xspace}
\usepackage{multirow}

\usepackage{amsmath}

\usepackage{supertabular}

\IEEEoverridecommandlockouts

\renewcommand{\textbf}[1]{\begingroup\bfseries\mathversion{bold}#1\endgroup}

\begin{document}

\title{Http-Burst: Improving HTTP Efficiency in the Era of Bandwidth Hungry Web Applications}

\author{
    \IEEEauthorblockN{Martin L\'evesque\\}
    \IEEEauthorblockA{School of Information Sciences\\
    University of Pittsburgh\\
    Pittsburgh, PA, USA
}
\thanks{Corresponding author: Martin L\'evesque, email: \texttt{levesque.martin@gmail.com}.}
}


\IEEEaftertitletext{\vspace*{-.7cm}
}

\maketitle
\begin{abstract}
The Hypertext Transfer Protocol (HTTP), a key building block of the World Wide Web, has succeeded to enable information exchange worldwide. Since its first version in 1996, HTTP/1.0, the average number of inlined objects and average total bytes per webpage have been increasing significantly for desktops and mobiles, from 1-10 objects in 1996 to more than 100 objects in June 2014. Even if the retrieving of inlined objects can be parallelized as a given Hypertext Markup Language (HTML) document is streamed, a maximum number of connections is allocated, and thus as the number of inlined objects increases, the overall webpage load duration grows, and the HTTP servers loading also gets higher. To overcome this issue, we propose a new HTTP method called BURST, which allows to retrieve the missing inlined objects of a webpage efficiently by requesting sets of web objects. We experimentally demonstrate the potential via a proof-of-concept demonstration, by comparing the regular HTTP to proposed HTTP-Burst using a virtual private server and real HTTP client and server over the Internet. The results indicate a latency reduction of webpage load duration compared to HTTP as high as 52 \% under the considered configurations.
\end{abstract}
\vspace{-.25cm}

\section{Introduction}

The Hypertext Transfer Protocol (HTTP) \cite{fielding1999hypertext}, which first version was established in 1996, is a key building block protocol in the Internet. HTTP works on a request-responce manner, where the GET, POST, PUT, and DELETE are the most commonly used methods. The most frequent method from an end-user standpoint is the GET method, which retrieves a web document from an HTTP server. A typical document type is the HyperText Markup Language (HTML), which contains HTML elements as well as reference to external objects, the so-called inlined objects. Each of those inlined objects are retrieved via successive GET requests, typically in parallel as the HTML document is obtained.

Several different optimization techniques have been proposed to improve the HTTP performance, from the client, network, and server viewpoints in order to minimize the request latency. To avoid fetching two times the same exact document, the end-user can store locally the document, which is possible by using a local cache. Proxy web caches are also widely used, where an intermediate node between the clients and server stores frequently used documents \cite{niclausse1997new, cao1999active, abrams1995caching}. Caching can also be done on the server side\cite{sivasubramanian2007analysis}, mainly to decrease the request computation duration especially when complex calculations or database queries are executed. Web caching on the server side is frequently done by saving content in memory for fast lookup \cite{davison2001web}. A typical performance metric used in web caching is the hit rate, which corresponds to the ratio of the number of requests using the cache over the number of requests.

Other improvement techniques focused on the exchanged HTTP messages. For instance, the HTTP data can be compressed to reduce the transmitted bytes \cite{liu2005http}. Further, it is worth noting that HTTP works by maintaining Transmission Control Protocol (TCP) connections. HTTP 1.1, as opposed to HTTP 1.0, introduced the persistent connections (HTTP Keep-alive), allowing to make multiple HTTP requests/responses over the lifetime of a given connection. This mechanism significantly reduces the overhead, as verified in \cite{nielsen1997network}. It thus improves the transport layer communications for the HTTP traffic. 

All of these techniques improve the request-response latency. It is worth nothing that the HTTP request-response latency performance highly depends on the number of inlined objects (e.g., images, fonts, etc.) to retrieve \cite{getall}. Since 1995 up to now, the number of inlined objects per webpage has significantly increased, as depicted in Fig. \cite{webOptimWebsite}, from few objects to more than 100 objects in 2014. In \cite{getall}, the authors measured this issue in the early HTTP version and proposed new methods, GETALL and GETLIST. GETALL, as opposed to the GET method, requests to retrieve a given document and its inlined objects for improved efficiency.

\begin{figure}[t]
\begin{center}
\includegraphics[width=.45
\textwidth]{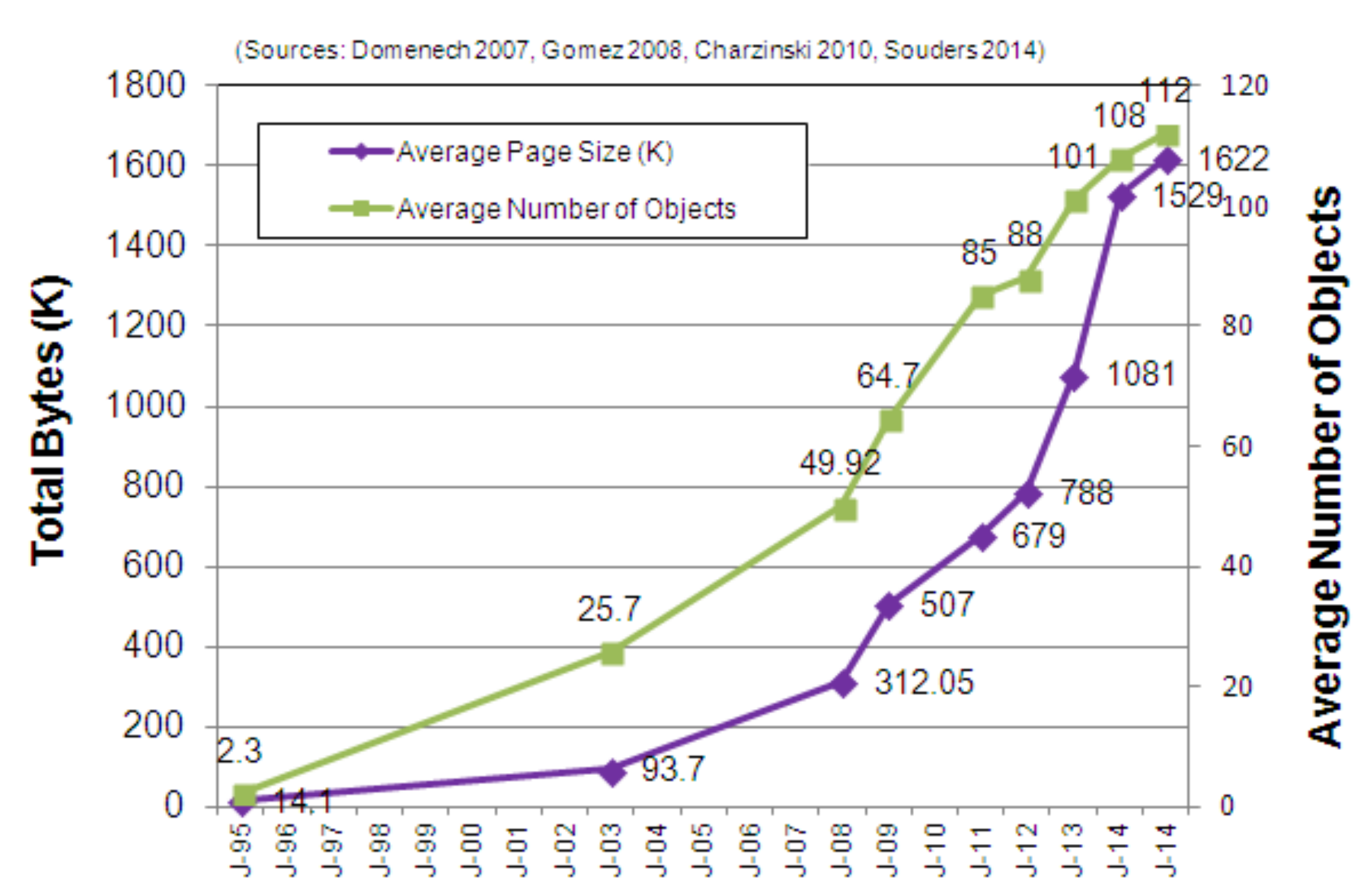}
\caption{Evolution of the average page size and number of objects from 1995 to 2014 \cite{webOptimWebsite}.}
\label{fig:growthAveragePageSizeFrom1995}
\end{center}
\end{figure}

In this paper, we propose the BURST method, which is somewhat a variant of the GETALL method proposed in \cite{getall}. However, rather than requesting the document and all inlined objects, the proposed extension first requests the document and then the missing inlined objects, and it allows to have multiple BURST requests to be sent in parallel to decrease latency.

This paper is structured as follows. In Section \ref{sec:trafficTrends}, we overview the HTTP traffic trends to motivate the proposed method. In Section \ref{sec:http}, we describe the overall HTTP protocol and propose an extension in Section \ref{sec:httpBurst}. In Section \ref{sec:experimentalResults}, we investigate experimentally the proposed method and compare it to the regular HTTP protocol. Section \ref{sec:conclusions} finally draws some conclusions.

\section{HTTP Traffic Trends}
\label{sec:trafficTrends}

\begin{figure}[!h]
\centering
\begin{subfigure}[b]{0.85\linewidth}
  \centering
  \includegraphics[width=1.0\textwidth]{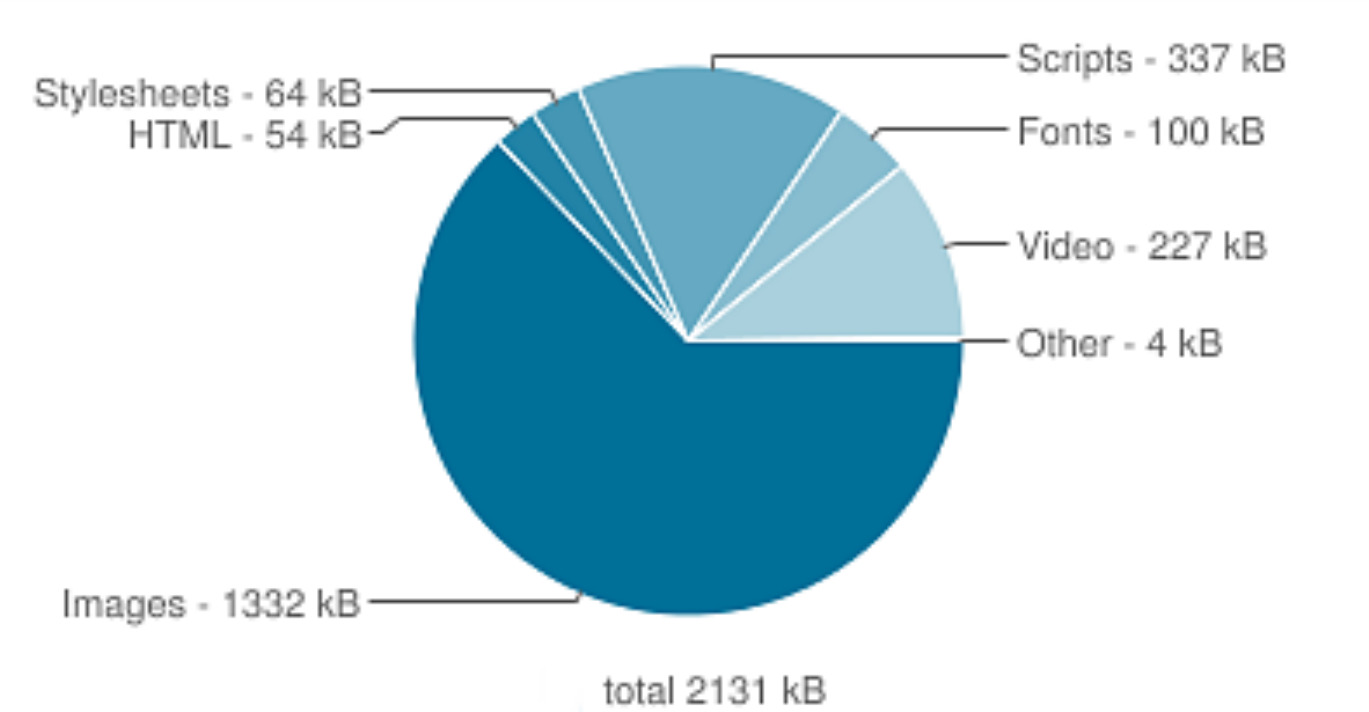}
  \caption{Desktop.}
  \label{fig:avgBytesContentTypeDesktop}
\end{subfigure}\\
\begin{subfigure}[b]{0.85\linewidth}
  \centering
  \includegraphics[width=1.0\textwidth]{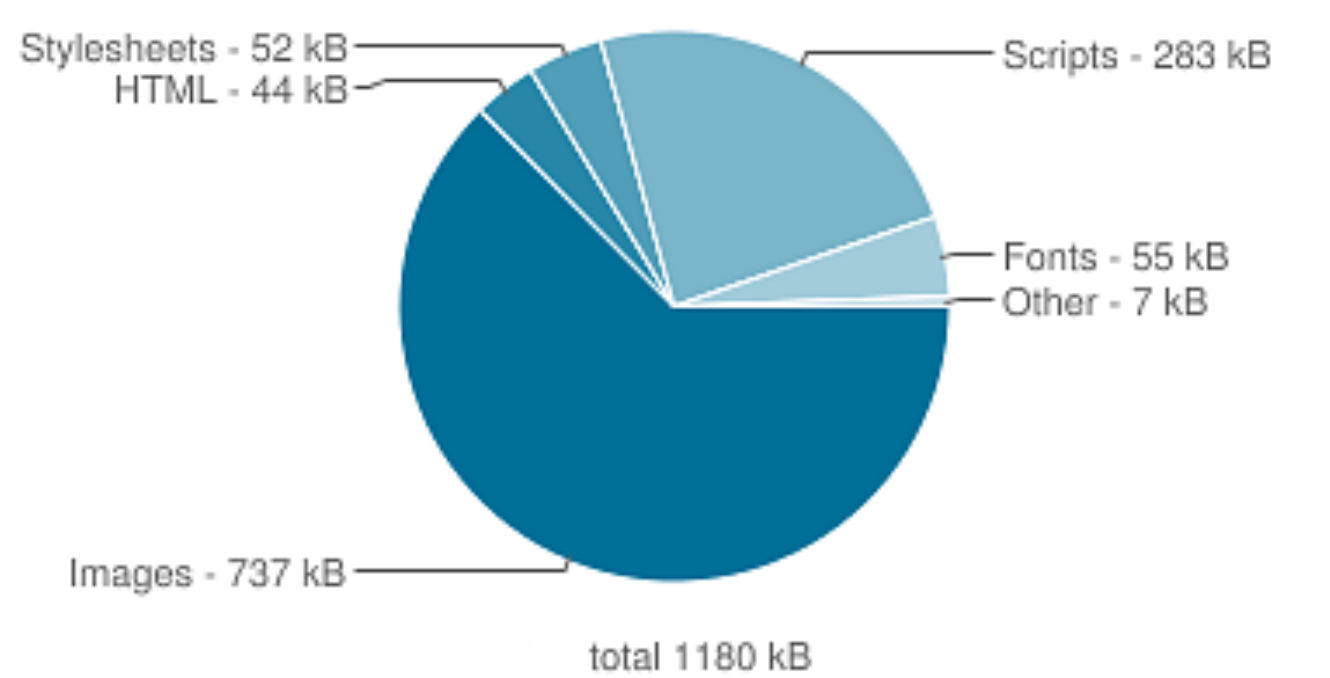}
  \caption{Mobile.}
  \label{fig:avgBytesContentTypeMobile}
\end{subfigure}\hfill
\caption{HTTP average object size per content type (June 2015) \cite{httpArchiveStats}.}
\label{fig:avgBytesContentType}
\end{figure}

	In this section, we overview the trends of the traffic generated by HTTP-based web applications. In 1996, \url{httparchive.org} was founded, aiming at keeping snapshots of the websites evolution by archiving the different versions of webpages over time. More interestingly for the scope of this paper, statistics can be generated of the HTTP-based Internet traffic \cite{httpArchiveStats}. We are interested to look at the traffic, in average, generated by a webpage. Fig. \ref{fig:avgBytesContentType} depicts the total average size in June 2015 of an accessed webpage, namely 2.1 MB (Fig. \ref{fig:avgBytesContentTypeDesktop})) from desktops and 1.18 MB (Fig. \ref{fig:avgBytesContentTypeMobile})) from mobile devices. Note that it does not take into account the scenario of a webpage where a given client has cached objects from previous requests, but illustrates the scenario where a user first access a given website/webpage without saved objects. Further, Fig. \ref{fig:avgBytesContentType} illustrates the average data size for the main downloaded object types while loading a page.

If we look more specifically on the webpages generated for desktops (Fig. \ref{fig:avgBytesContentTypeDesktop})), the two average largest content type sizes are originating from images and scripts (e.g., Javascript files). On the mobile counterpart (Fig. \ref{fig:avgBytesContentTypeMobile})), the average content sizes are reduced, mainly according to the images, scripts, and fonts sizes.

One could however wonder how did the total webpage size has evolved since 1995 and beyond. In Fig. \ref{fig:avgBytesContentType}, one can observe that the average webpage size has been growing since 1995 until now, and will more likely continue to grow given this trend. More specifically, the average webpage size has been growing significant since 2010, perhaps related to the economic growth that followed the economic crisis of 2008.

\begin{figure}[t]
\begin{center}
\includegraphics[width=.45
\textwidth]{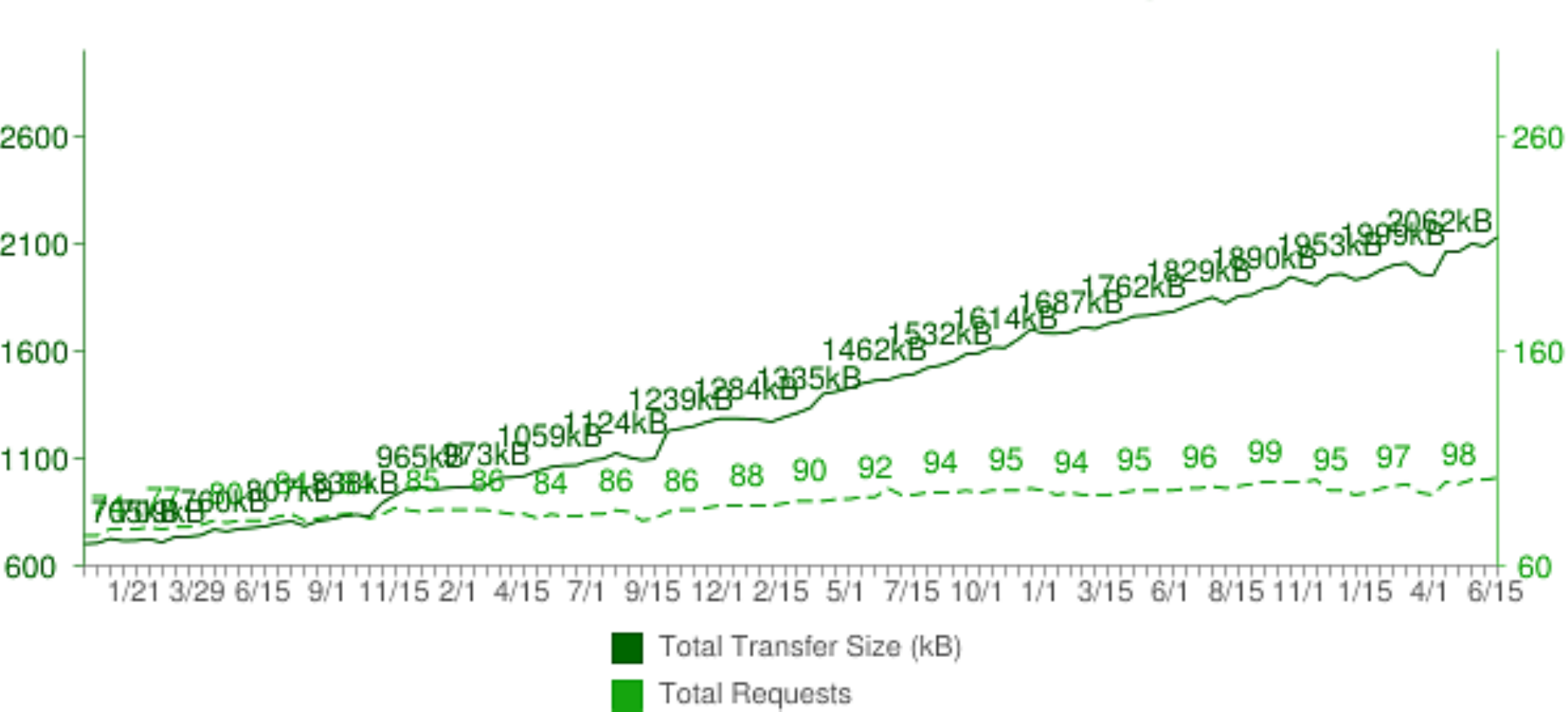}
\caption{Average page size and number of objects from 2010 to 2015 \cite{httpArchiveStats}.}
\label{fig:growthAveragePageSizeFrom2014}
\end{center}
\end{figure}

\begin{figure}[!h]
\centering
\begin{subfigure}[b]{0.85\linewidth}
  \centering
  \includegraphics[width=1.0\textwidth]{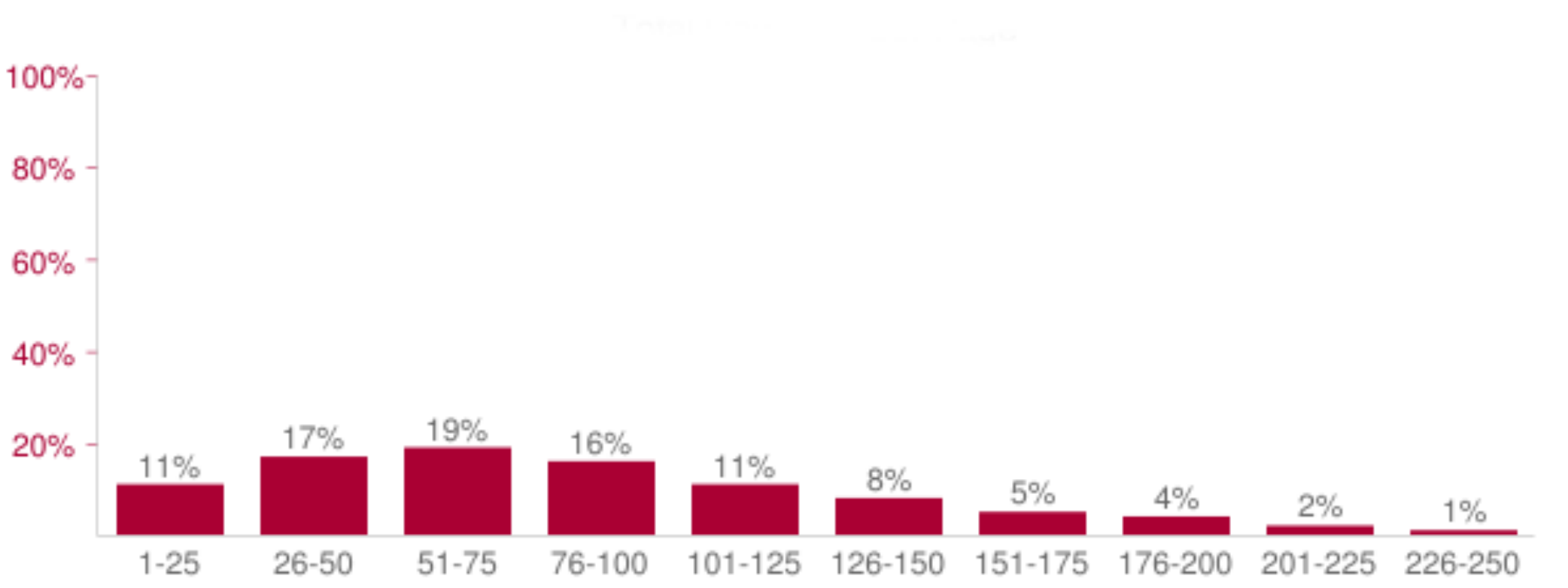}
  \caption{HTTP requests per page.}
  \label{fig:requestsPerPage}
\end{subfigure}\hfill
\begin{subfigure}[b]{0.85\linewidth}
  \centering
  \includegraphics[width=1.0\textwidth]{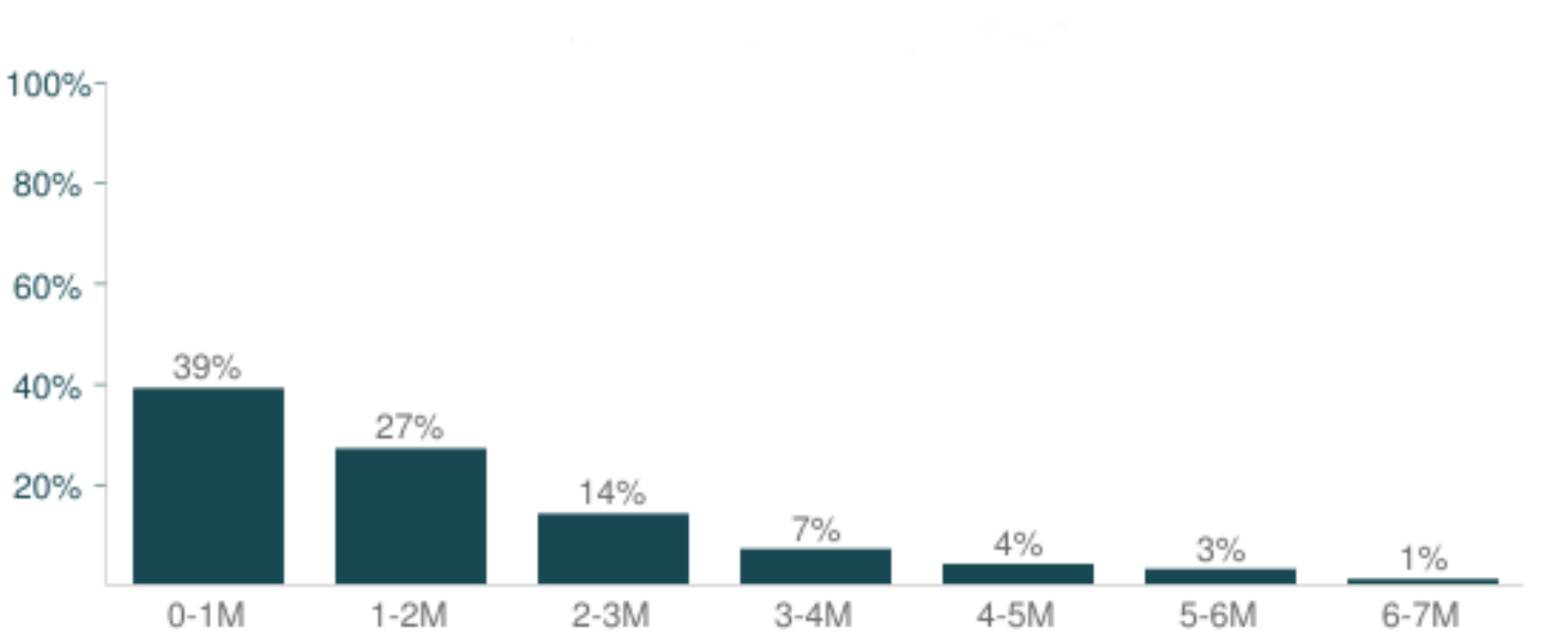}
  \caption{Transmitted MB per page (including all objects).}
  \label{fig:txPerPage}
\end{subfigure}\hfill
\caption{HTTP requests and transmitted KB per page (June 2015) \cite{httpArchiveStats}.}
\label{fig:requestsAndTXPerPage}
\end{figure}

It is worth noting that an HTTP page load, which we will briefly overwiew in the next section, basically consists of an HTML document request, followed by successive object file requests for images, scripts, etc. Fig. \ref{fig:requestsAndTXPerPage} depicts the averages total number of requests and total bytes a webpage is creating. Approximately 75 \% of webpage loads require between 1 and 125 HTTP requests, which is quite significant, given that each given HTTP request further generates overhead from the IP and TCP layers, as well as increasing server loading.

\begin{figure}[!h]
\centering
\begin{subfigure}[b]{0.85\linewidth}
  \centering
  \includegraphics[width=1.0\textwidth]{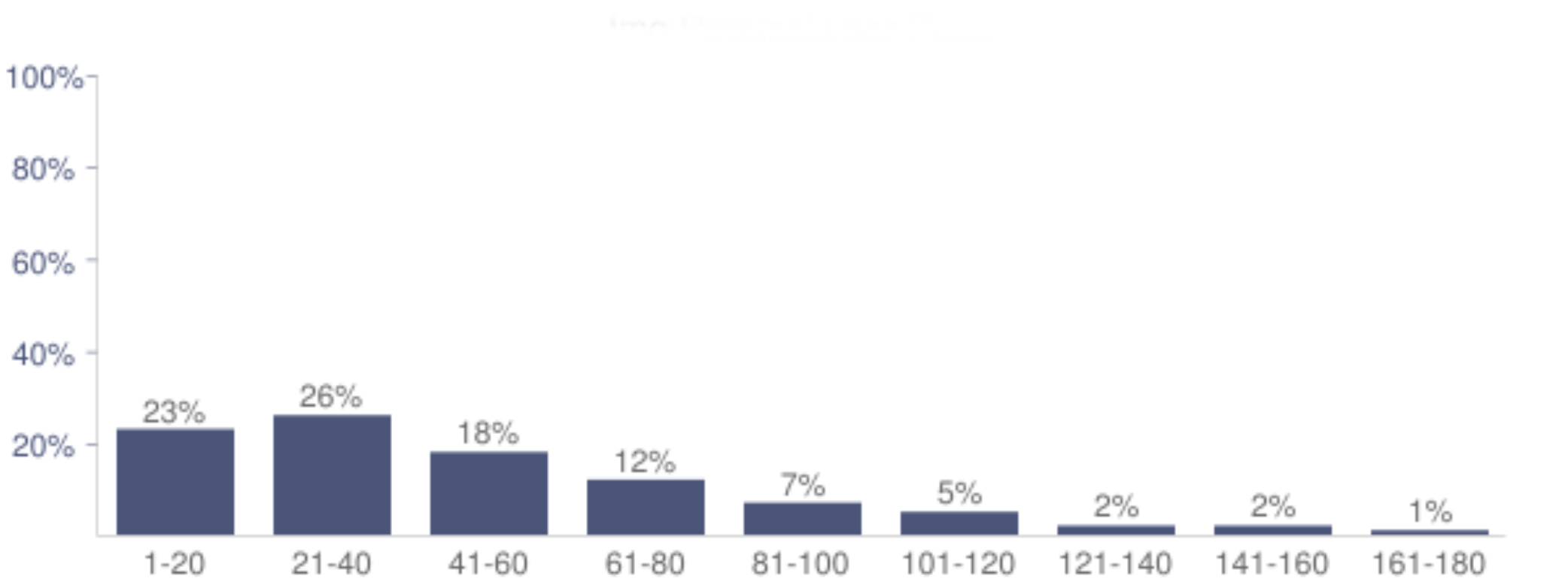}
  \caption{Images per page.}
  \label{fig:imagesPerPage}
\end{subfigure}\hfill
\begin{subfigure}[b]{0.85\linewidth}
  \centering
  \includegraphics[width=1.0\textwidth]{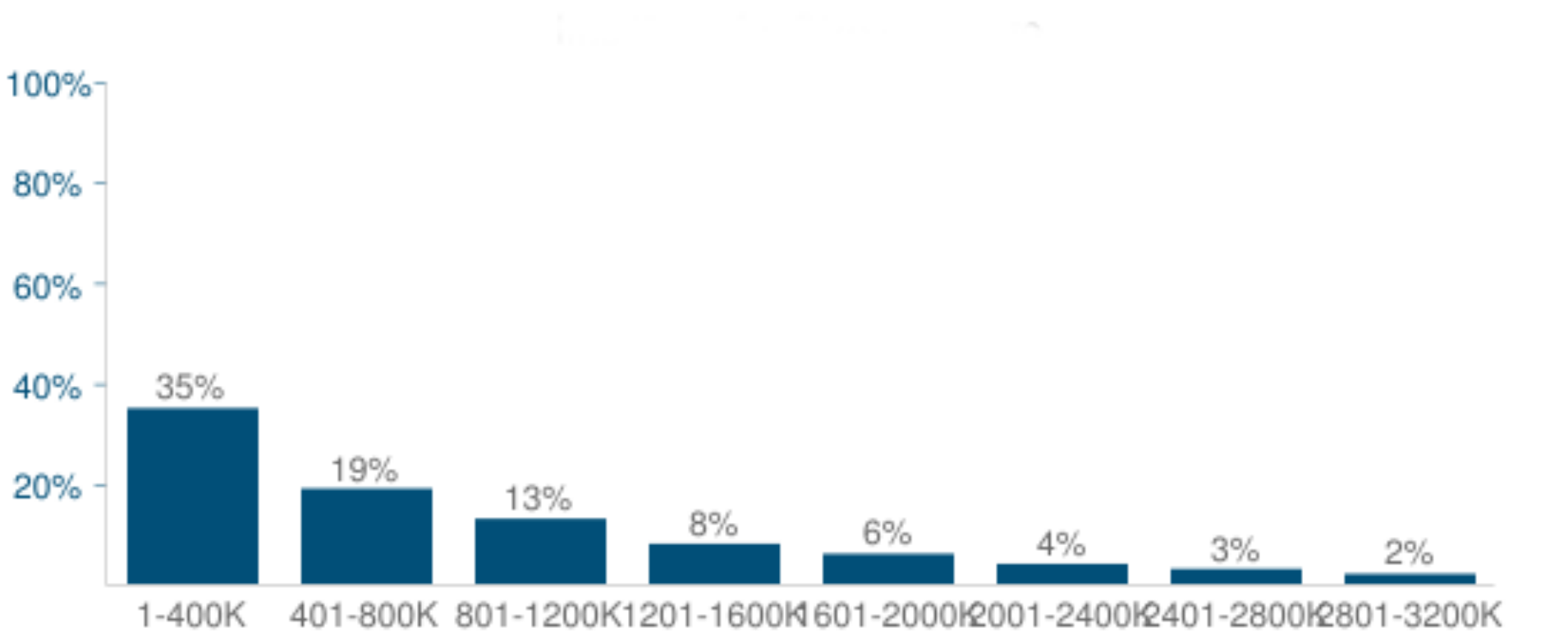}
  \caption{Images size per page (including all images).}
  \label{fig:imagesSizePerPage}
\end{subfigure}\hfill
\caption{Number and total size of images per page (June 2015) \cite{httpArchiveStats}.}
\label{fig:imagesAndTXPerPage}
\end{figure}

If we look in Fig. \ref{fig:imagesSizePerPage}) at the content type generating the largest amount of traffic, the images, we observe that 85 \% of webpages contain between 1 and 80 images.

If we multiply the ratios to the number of requests in Fig. \ref{fig:requestsPerPage}), approximately 91 requests are generated per webpage load. In the following, we overview HTTP to outline the consequence of having such a growing amount of requests. 

\section{Hypertext Transfer Protocol (HTTP)}
\label{sec:http}

\begin{figure}
\begin{center}
\includegraphics[width=.40
\textwidth]{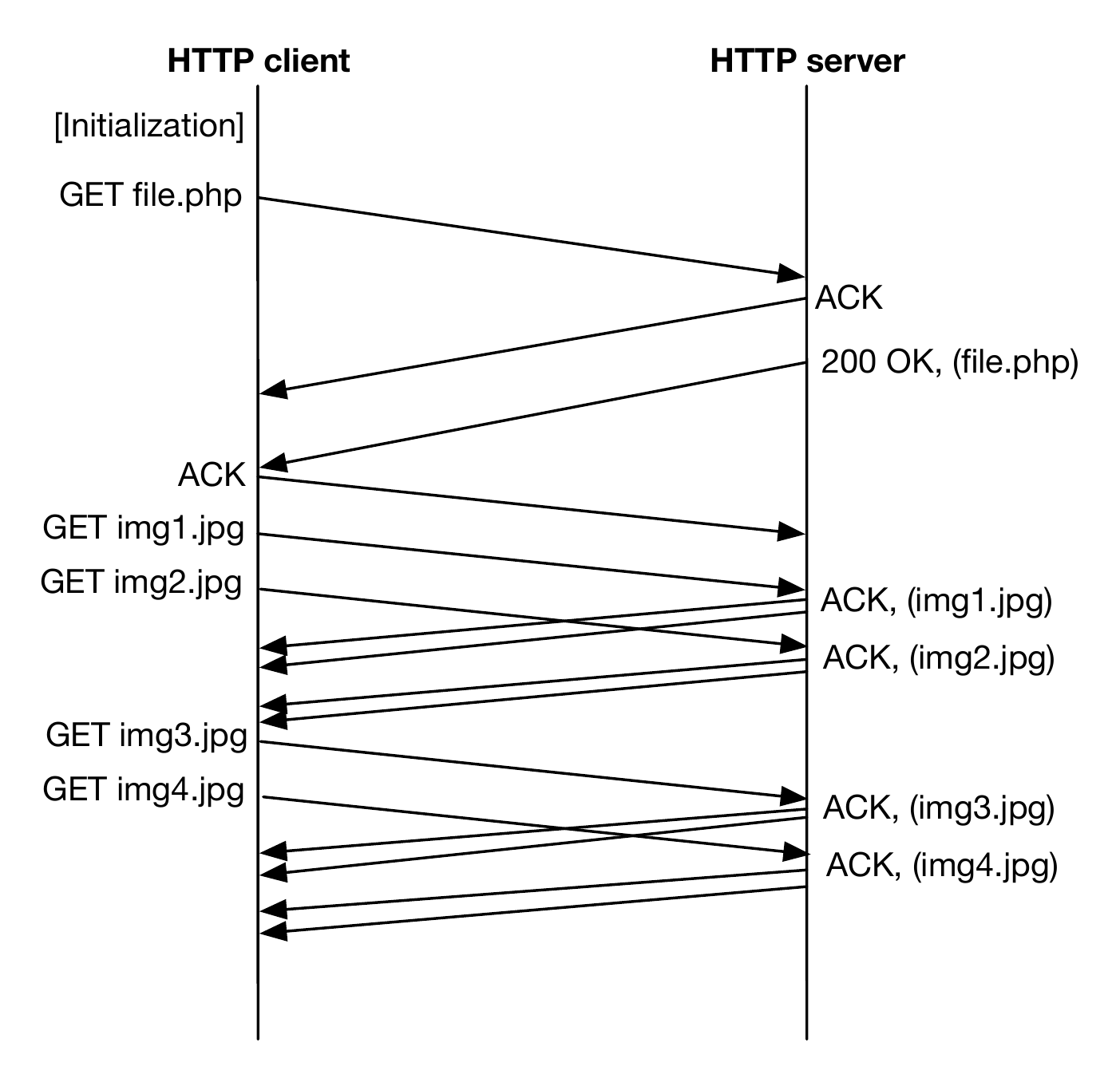}
\caption{HTTP example execution of one HTML file containing four images using the GET method ($\mathcal{C} = 2$).}
\label{fig:http}
\end{center}
\end{figure}

The Hypertext Transfer Protocol (HTTP) is an application protocol for distributed and collaborative information systems, which is one of the most widely used protocol in the Internet \cite{fielding1999hypertext}. HTTP has a variety of request methods, including GET, HEAD, POST, and DELETE. The GET method is the most widely used method, which role is to retrieve a specified resource, while the POST method is used to for instance in web forms to submit data entities. Fig. \ref{fig:http} depicts an HTTP GET example of a script request (file.php) containing four images. In the example, a maximum of two files can be transferred in parallel. Note that the number of parallel connections vary depending on the web browser, from 6 in Chrome and Firefox up to 13 in Internet Explorer 11 \footnote{Browserscope web browsers profiling: \url{http://www.browserscope.org/?category=network}.}. For each given request, TCP acknowledgement (ACK) packets are generated for confirmation. Therefore, as the number of objects in requested webpages increases, the experienced overhead grows.

\subsection{Maximum Efficiency}

As we illustrated in the previous section, the number of objects per webpage is significantly increasing over time since 1995. Let us consider $N$ objects (scripts, images, videos, etc.) included in a given webpage. Each object size $i$ corresponds to $L_i$ (in bytes) such that $i \in \{1, 2, ..., N\}$. An HTTP request for a given object $i$ thus totals the following minimum amount of bytes:

\begin{equation}
R_i = 2 \cdot (IP + TCP + HTTP) + ACK + L_i,
\end{equation}
where $IP$ and $TCP$ corresponds to the Internet Protocol (IP) and Transmission Control Protocol (TCP) header sizes, $HTTP$ the size of the HTTP request excluding  the payload of $L_i$ bytes, and $ACK = IP + TCP$ corresponds to the TCP ACK packet generated from the HTTP server to a given client. For instance, the IPv4 and TCP headers correspond to 20 bytes each. $IP + TCP + HTTP$ is counted twice due to the request and response messages. Note that, on the top of this, the link layer (e.g., Ethernet) also adds extra overhead which we do not take into account for improved readability. Therefore, the maximum efficiency is given by:

\begin{equation}
\frac{\sum_{i = 1}^N L_i}{\sum_{i = 1}^N R_i}.
\end{equation}

Thus, as the number of objects increases, the overhead data size increases. For instance, if the average object size corresponds to 200 bytes with $N = 3$, then the maximum efficiency is $\frac{3 \cdot 200}{3 \cdot 200 + 3 \cdot 120} = 62.5\ \%$.

\subsection{Delay}

In terms of delay, the number of objects has a major impact on the webpage load time, which corresponds to:

\begin{equation}
D_{max} = \sum_{i = 1}^N D_{c \rightarrow s}^i + p_{i, s} + D_{s \rightarrow c}^i,
\end{equation}
where $D_{c \rightarrow s}^i$ and $D_{s \rightarrow c}^i$ correspond to the delay between the client and server, and inversely. $p_{i, s}$ corresponds to the processing delay at the server to process object $i$. Note that this delay is correct if a single connection is used, which is usually not the case. If the browser uses $\mathcal{C}$ connections, then the delay $D$ corresponds to:

\begin{equation}
D = \frac{D_{max}}{\mathcal{C}}.
\end{equation}

It is clear that the webpage load time $D$ grows as the number of objects $N$ increases. To overcome this shortcoming, we propose HTTP-Burst in the following section.

\section{HTTP-Burst}
\label{sec:httpBurst}

\begin{figure}
\begin{center}
\includegraphics[width=.40
\textwidth]{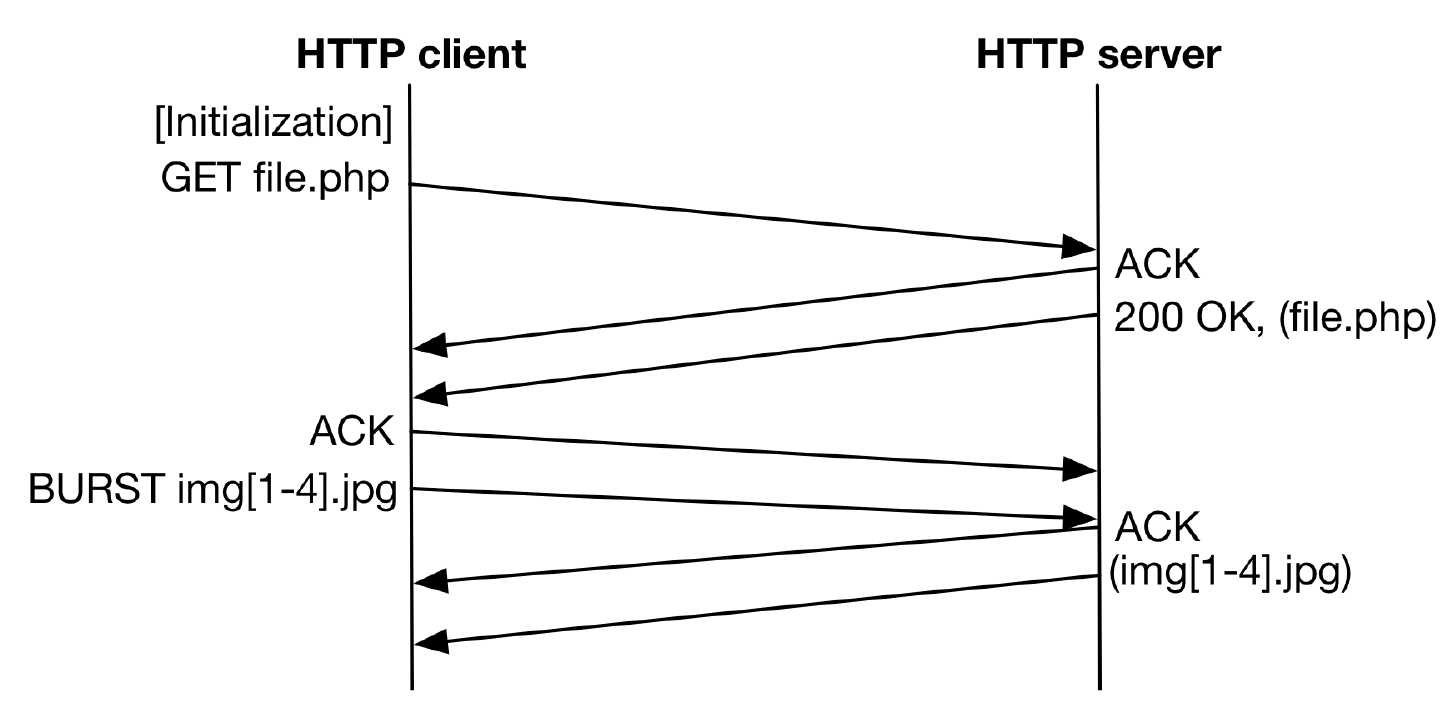}
\caption{HTTP-Burst example execution of one HTML file containing four images using the BURST method ($\mathcal{C} = 1$).}
\label{fig:httpBurst}
\end{center}
\end{figure}

\begin{figure}
\begin{center}
\includegraphics[width=.40
\textwidth]{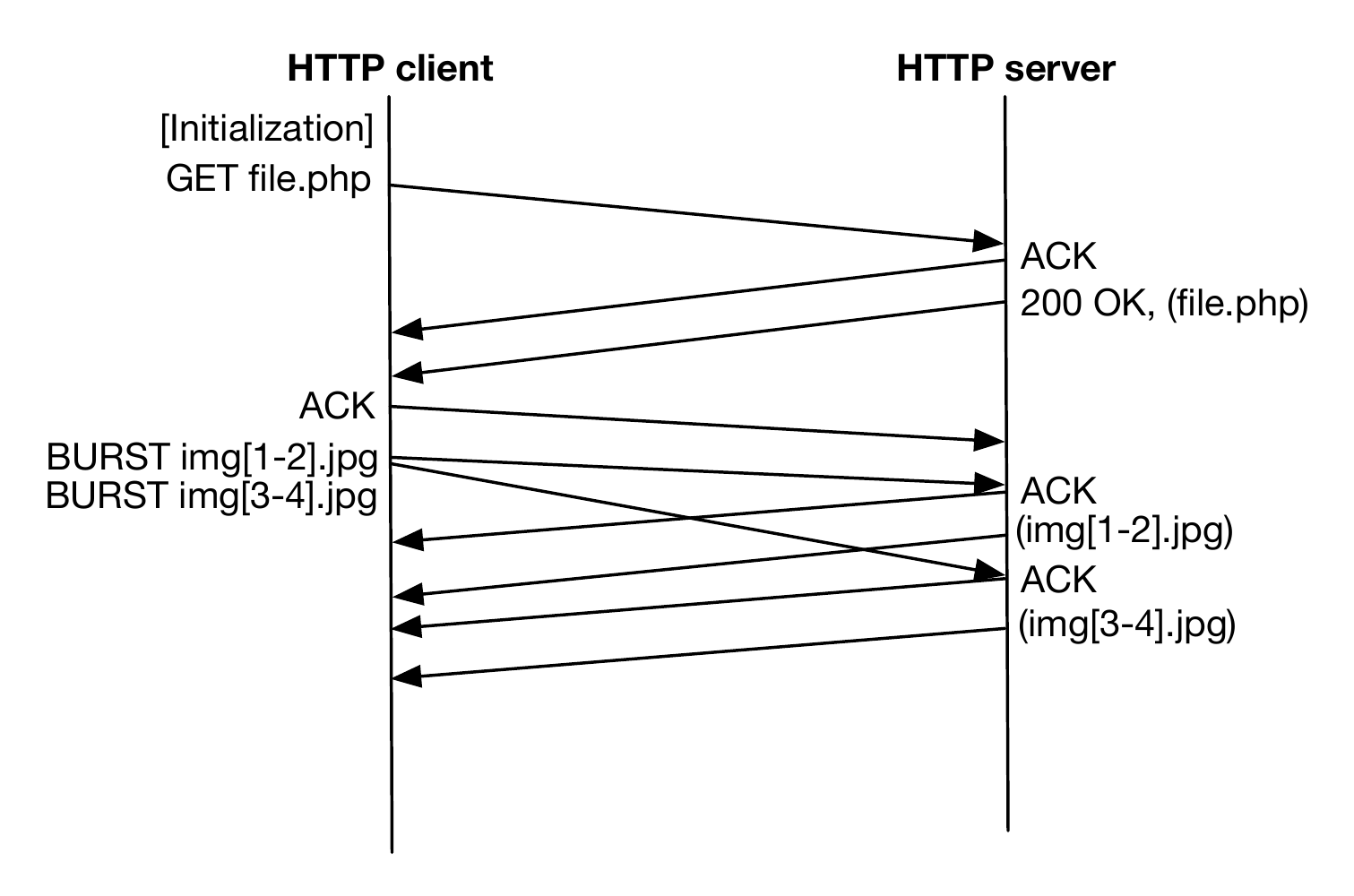}
\caption{HTTP-Burst example execution of one HTML file containing four images using the BURST method ($\mathcal{C} = 2$).}
\label{fig:httpBurstC2}
\end{center}
\end{figure}

For improved efficiency, we propose to transfer all missing objects, not stored in a local cache, in a single request as a burst using a new BURST method, as illustrated in Fig \ref{fig:httpBurst}. In such a scheme, when the GET method has been completed and the HTTP client has completely received the HTML content, one or many BURST requests are sent each containing a set of objects to retrieve from the HTTP server, for example to retrieve four images, img1.jpg, ..., img4.jpg with a single connection ($\mathcal{C} = 1$) as in Fig \ref{fig:httpBurst}. Then, the HTTP server transmits all $N$ requested files in $\mathcal{C}$ concatenated HTTP messages. As we can see, a significantly lower number of requests is being exchanged between the client and server, compared to using successive GET requests (Fig \ref{fig:http}). Fig. \ref{fig:httpBurstC2} depicts the same example, but with two connections, $\mathcal{C} = 2$. When two connections are used, the two BURST requests are sent to retrieve the object files. Therefore, each BURST requests for two objects.

\subsection{Maximum Efficiency}

To compare with HTTP, the maximum efficiency using the BURST method is given by:

\begin{equation}
\frac{\sum_{i = 1}^N L_i}{\mathcal{C} \cdot (2 \cdot (IP + TCP + HTTP) + ACK) + \sum_{i = 1}^N L_i}.
\end{equation}

\begin{figure}
\begin{center}
\includegraphics[width=.45
\textwidth]{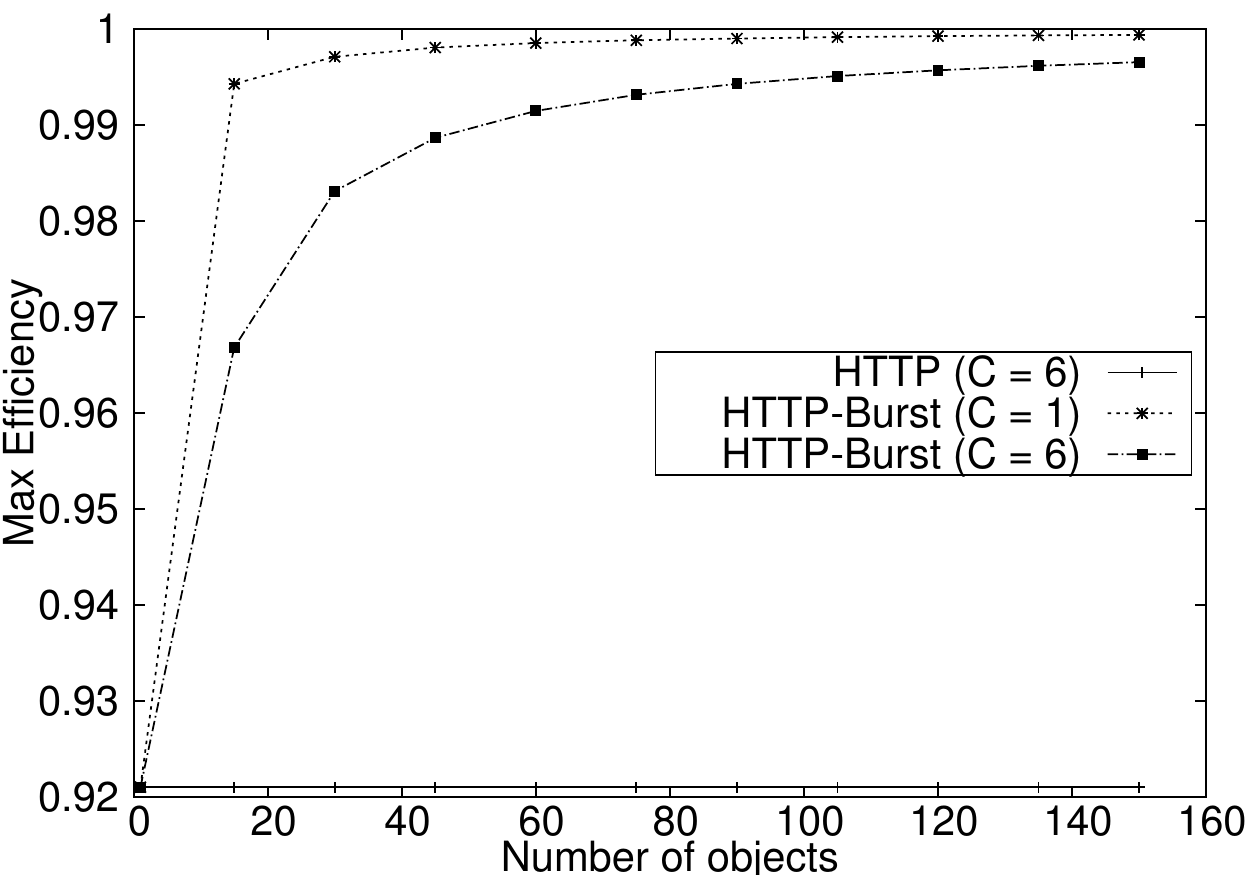}
\caption{Maximum efficiency, $L_i = 1400, \forall i$.}
\label{fig:maxEfficiency}
\end{center}
\end{figure}

Thus, the efficiency is clearly improved, as the TCP/IP overheads are counted maximally $\mathcal{C}$ times instead of $N$. Fig. \ref{fig:maxEfficiency} depicts the maximum efficiency with a fixed payload of 1400 bytes. The efficiency is similar for both only when the number of objects corresponds to $N \le \mathcal{C}$.

\subsection{Delay}

In terms of delay, it is definitely improved since few two-way communications exchanges are done:

\begin{equation}
D = max D_{c \rightarrow s}^{i} + p_{i, s} + D_{s \rightarrow c}^{i},
\end{equation}
where $c$ is the BURST connection identifier, $i = 1, 2, ..., \mathcal{C}$.

Further, as the number of requests is reduced, processing only $\mathcal{C}$ HTTP request instead of $N$ definitely reduces the server load.

\section{Experimental Results}
\label{sec:experimentalResults}

In this section, we experimentally demonstrate the potential of the proposed BURST method for HTTP. We made a proof-of-concept (POC), using HTTP GET requests. The source code we used is available online\footnote{POC source code: \url{https://github.com/martinlevesque/http-burst}.}. The scenario we investigate is as follows. We configured a virtual private server (VPS) with lighttpd\footnote{lighttpd: \url{http://www.lighttpd.net/}.}, which is a light and efficient HTTP server. The VPS has 1 GB random access memory (RAM), and does not experience any external requests except the ones received for this given experience. The experiment we consider consists of measuring the total request duration of retrieving an HTML document and its embedded objects. The HTML document contains a font of 44 KB, a CSS file of 120 KB, a javascript (jquery) file of 84 KB, and a variable number of images ($N$). Each request originates from a client located in Pittsburgh, PA, USA, and the VPS server is located in Dallas, Texas, USA. 

\begin{figure}
\begin{center}
\includegraphics[width=.45
\textwidth]{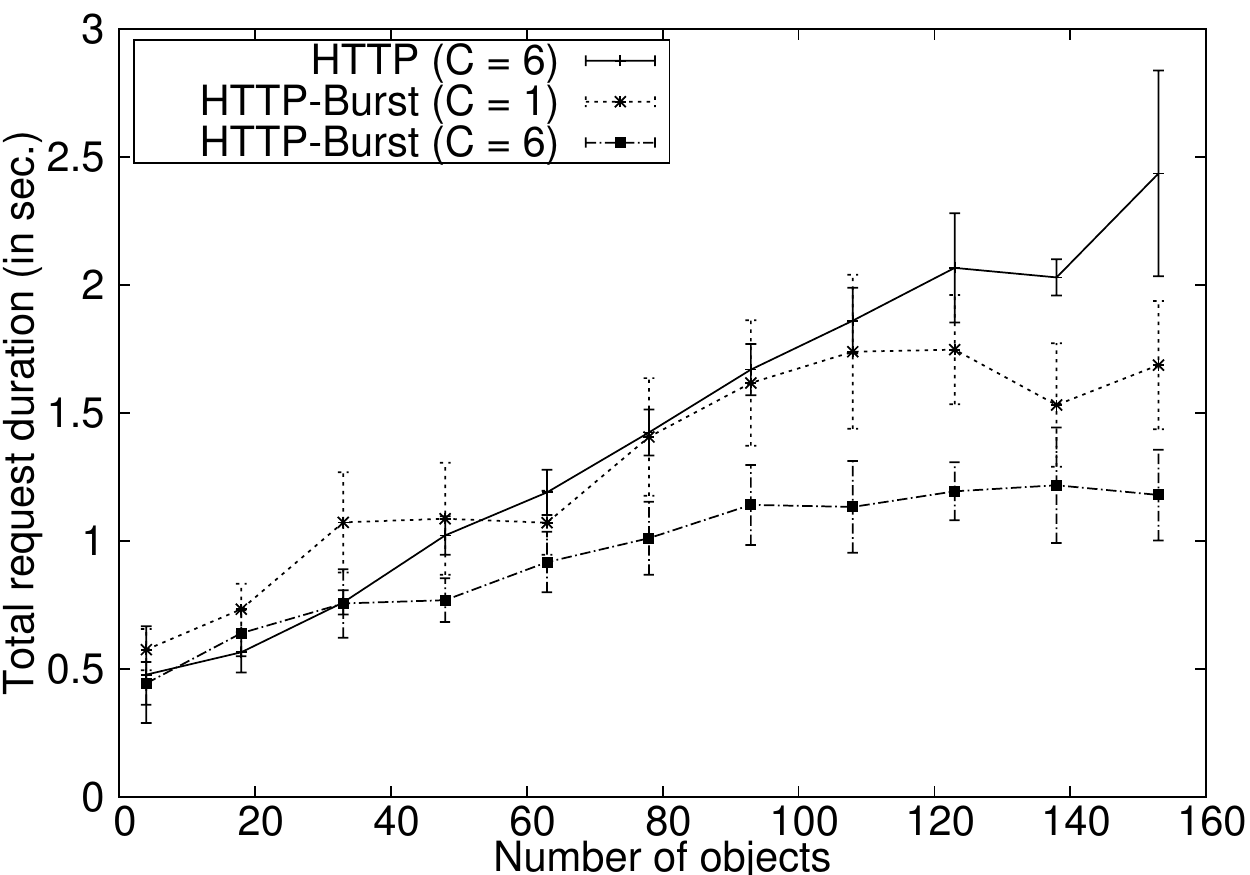}
\caption{Experimental measurements of the total request duration.}
\label{fig:result}
\end{center}
\end{figure}

Fig. \ref{fig:result} compares HTTP using GET requests and the proposed HTTP-Burst protocol, with $\mathcal{C} = \{1, 6\}$, which is the maximum allowed number of parallel connections. We vary the number of images, where the number of objects corresponds to the number of images, CSS, javascript, and font files. For each given number of images, we process the request 10 times, and we record the mean and standard deviation. On Fig. \ref{fig:result}, we thus plot the mean and $\pm$ the standard deviation to illustrate the fluctuations. For each request, the total request duration is recorded, which corresponds to the time duration from the HTML request sending to the reception of all retrieved objects. We observe that for HTTP, the request duration grows from 0.5 second to approximately 2.5 seconds. With HTTP-Burst, however, similar request duration are experienced for low number of objects. For larger number of objects (more than 30), HTTP-Burst outperforms the performance of the regular HTTP protocol. For instance, with HTTP-Burst and $\mathcal{C} = 6$, even with 150 objects, the request duration is approximately as low as 1.2 seconds, compared to 2.5 seconds with HTTP GET requests. With HTTP-Burst and $\mathcal{C} = 1$, the total request duration is worse compared with $\mathcal{C} = 6$ for low number of objects. However, as the number of objects increases, HTTP-Burst with $\mathcal{C} = 1$ still outperforms the regular HTTP protocol using GET requests. In terms of percentage improvement with the largest number of objects, 150, and HTTP-Burst with $\mathcal{C} = 1$ and $\mathcal{C} = 6$, improvements of $\frac{2.5-1.75}{2.5} \approx 30 \%$ and $\frac{2.5-1.2}{2.5} \approx 52 \%$ are obtained, respectively.

\section{Conclusions}
\label{sec:conclusions}

In this paper, we first overviewed the HTTP traffic trends. The average webpage size and number of objects per webpage have continuously increased since 1995. In 2015, the average number of objects per webpage for desktops has reached 200 objects. Even if objects can be retrieved with parallel TCP connections, the increasing number of objects causes significant overhead both from the network and server viewpoints. To overcome this issue, we proposed HTTP-Burst, which, instead of sending HTTP requests for all objects, retrieves webpage objects as bursts via the proposed BURST method, where objects are grouped for improved efficiency. We experimentally investigated the proposed scheme by comparing it to using sucessive HTTP GET requests. The results shown a reduction of as much as 52 \% of the total request duration under the considered configurations, thus illustrating the potential of the proposed protocol extension. In future work, we will look at the principal overhead causes and experiment the scheme under different conditions.

\bibliographystyle{IEEEtran}
\bibliography{httpBurst}
  
\end{document}